\documentclass[a4paper,11pt]{article}

\usepackage{pos_MOU}
\usepackage{caption}
\usepackage{subcaption}
\usepackage{graphicx}
\usepackage{booktabs}
\usepackage{multirow}

\title{H.E.S.S.~realtime follow-ups of IceCube high-energy neutrino alerts}

\ShortTitle{H.E.S.S.~realtime follow-ups of IceCube high-energy neutrino alerts}

\author*[1]{Federica Bradascio}
\author[2]{Halim Ashkar} 
\author[3]{Jowita Borowska} 
\author[4]{Jean Damascene Mbarubucyeye}
\author[5]{Enzo Oukacha}
\author[1]{Fabian Sch\"{u}ssler}
\author[6]{Hiromasa Suzuki}
\author[7]{Alicja Wierzcholska}

\affiliation[1]{IRFU, CEA, Universit\'e Paris-Saclay, F-91191 Gif-sur-Yvette, France}
\affiliation[2]{Laboratoire Leprince-Ringuet, École Polytechnique, CNRS, Institut Polytechnique de Paris, F-91128 Palaiseau, France}
\affiliation[3]{Institut f\"{u}r Physik, Humboldt-Universität zu Berlin, D-12489 Berlin, Germany}
\affiliation[4]{Deutsches Elektronen-Synchrotron (DESY), Platanenallee 6, 15738 Zeuthen, Germany}
\affiliation[5]{Université Paris Cité, CNRS/IN2P3, AstroParticule et Cosmologie (APC), Paris F-75013, France}
\affiliation[6]{Department of Physics, Konan University, 8-9-1 Okamoto, Higashinada-ku, Kobe, Hyogo, 658‐8501, Japan}
\affiliation[7]{Instytut Fizyki Jadrowej PAN, ul. Radzikowskiego 152, 31-342 Kraków, Poland}

\forColl{H.E.S.S.~and IceCube} 
\emailAdd{federica.bradascio@cea.fr}

\newcommand{\hess}[0]{H.E.S.S.}
\newcommand{\fermi}[0]{\textit{Fermi}-LAT}
\newcommand{\swift}[0]{\textit{Swift}-XRT}

\newcommand{\icrevA}[1]{{#1}} %
\newcommand{\icrevB}[1]{{#1}} %
\newcommand{\hessrevA}[1]{{#1}} 
\newcommand{\hessrevB}[1]{{#1}} 
\newcommand{\hessrev}[1]{{#1}} 

\abstract{

The evidence for multi-messenger photon and neutrino emission from the blazar TXS 0506+056 has demonstrated the importance of realtime follow-up of neutrino events by various ground- and space-based facilities. The effort of \hess{} and other experiments in coordinating observations to obtain \hessrevA{quasi-}simultaneous multiwavelength flux and spectrum measurements has been critical in measuring the chance coincidence with the high-energy neutrino event IC-170922A and constraining theoretical models. For about a decade, the \hess{} transient program has included a search for gamma-ray emission associated with high-energy neutrino alerts, looking for gamma-ray activity from known sources and newly detected emitters consistent with the neutrino location. In this contribution, we present an overview of follow-up activities for realtime neutrino alerts with H.E.S.S. in 2021 and 2022. \icrevB{Our analysis includes both public IceCube neutrino alerts and alerts exchanged as part of a joint \hess{}--IceCube program.} 
 \hessrevA{We focus on interesting coincidences observed with gamma-ray sources, particularly highlighting the significant detection of PKS 0625-35, an AGN previously detected by H.E.S.S., and three IceCube neutrinos.} }

\ConferenceLogo{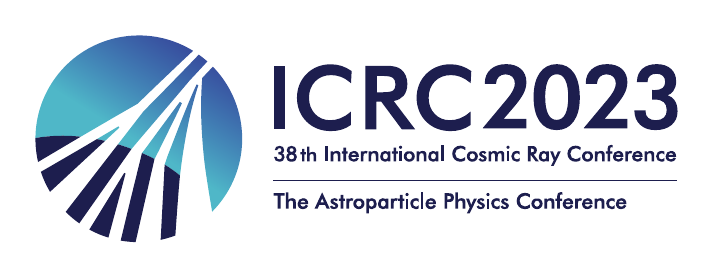}

\FullConference{The 38th International Cosmic Ray Conference (ICRC2023)\\ 26 July -- 3 August, 2023\\ Nagoya, Japan}

\let\OLDthebibliography\thebibliography
\renewcommand\thebibliography[1]{
  \OLDthebibliography{#1}
  \setlength{\parskip}{1pt}
  \setlength{\itemsep}{2pt plus 0.5ex}
}

\begin{document}

\maketitle
\section{Introduction}

The IceCube Neutrino Observatory in Antarctica \icrevA{discovered an astrophysical flux of high-energy neutrinos in 2013}~\cite{IceCube13}, prompting a search for their sources. Despite the strong evidence of TeV neutrino emissions from the active galaxy NGC~1068 \cite{IceCube2022_NGC1068}, the majority of the sources remain unidentified due to the limited angular resolution of neutrino detectors, the background noise from atmospheric neutrinos, and the low signal statistics ($\sim$10/year). To overcome these challenges, a multi-messenger approach combining observations \hessrevA{in different} wavelengths and astrophysical messengers is employed. Very-high-energy (VHE, $E>100$~GeV) gamma rays are particularly \hessrevA{promising} in identifying neutrino sources, as they are produced alongside neutrinos through \hessrevA{cosmic-ray} interactions. Unlike neutrinos, gamma rays can be absorbed or scattered during their journey through the source region and extragalactic space\hessrevA{. However,} their detection or non-detection consistent with high-energy neutrinos provides valuable information about the distances and environments of the sources. Notably, the identification of the flaring gamma-ray blazar TXS~0506+056 \hessrevA{coincident} with a single, highly energetic neutrino in 2017 demonstrated the effectiveness of the multi-messenger approach~\cite{IceCube2018}. In this contribution, the follow-up strategy of the \hess{} imaging atmospheric Cherenkov Telescope (IACT) \hessrev{array} to real-time neutrino alerts is presented, along with a summary of results from November 2021 to the end of 2022. \icrevB{Our analysis includes both public IceCube neutrino alerts and alerts exchanged as part of a joint \hess{}-IceCube program.}

\section{H.E.S.S. neutrino follow-up program}

\hessrevA{\hess{} is an array consisting of one 28 m and four 12 m IACTs, located in the Khomas Highland of Namibia, at an elevation of 1800~m above sea level, with a field-of-view (FoV) of $5^{\circ}$ in diameter~\cite{hess_aharonian_2006}}. 
The H.E.S.S. observatory has been engaged in a neutrino Target of Opportunity (ToO) program since 2012, aimed at investigating potential correlations between neutrino and VHE gamma-ray emission. Initially triggered by real-time alerts from the ANTARES neutrino telescope, \hess{} later joined the IceCube Realtime Alert Stream~\cite{realtime_ic} and \icrevB{IceCube's} Gamma-ray Follow-Up (GFU) program~\cite{gfu}. \hessrevA{\hess{} dedicates approximately 20 hours of observation time per year to deep investigations of intriguing neutrino events, with the goal of detecting  VHE gamma-ray emissions. In case of detection, longer and deeper follow-up observations are triggered to perform source characterization~\cite{hess_ic_followup_first}.}

This contribution presents the follow-up of six neutrino alerts received through IceCube's singlet alert stream and the GFU program, with singlet alerts comprising high-energy single events of likely astrophysical origin, categorized into \textit{gold} and \textit{bronze} alerts based on their signalness\footnote{Probability that this is an astrophysical signal relative to backgrounds, assuming the best-fit diffuse muon neutrino astrophysical power-law flux $E^{-2.19}$.} probability~\cite{ic_alerts}.
\hessrev{Events with an average signalness value above 50\%  are flagged and distributed as \textit{gold} alerts, while those with an average signalness value between 30\% and 50\% are categorized as \textit{bronze} alerts.} 
\icrevB{These events are broadcast by IceCube in realtime with a typical latency of $\sim$30~s, and have a localization uncertainty of $\sim1^{\circ}$, matching a typical IACT field of view of $3.5-5^{\circ}$.}
The GFU alerts, emitted when clusters of neutrino events around a priori defined list of sources surpass predefined significance thresholds, are accessible to H.E.S.S. through a Memorandum of Understanding (MoU), allowing for fully automated follow-up observations via a VOEvent-based alert stream~\cite{hess_transients}. \hessrevA{Without prior identification of promising source candidates (e.g., using the \fermi{} and IACT catalogs), the searches typically encompass the entire region of interest (ROI) defined by the uncertainty in neutrino localization. If the external conditions permit, such as receiving the alert during a dark night when the source is visible and the weather is favorable, an automatic re-pointing procedure is initiated by the telescopes, enabling immediate observation of the alert. Otherwise, observations usually occur within the next few days, once these conditions are met. As a result, the response time varies for each individual alert, with the fastest recorded at less than 70 seconds.}

The \hess{} data presented in this contribution were analyzed using the {method} described in \cite{hess_denaurois_2009} with \hessrevA{estimated} gamma-hadron separation and event selection cuts. The background was determined using the standard ``{reflected} background" technique \cite{hess_berge_2007}. The results were validated by a \hessrevA{cross-check} analysis which uses an independent event calibration and reconstruction~\cite{hess_parsons_2014}.
The flux upper limit calculation is performed using the Rolke method and assuming a power-law spectrum with a photon index of \hessrevA{2.0} \cite{hess_rolke_2005}. 
The minimum energy is chosen as the energy where the effective area reaches $10\%$ of its maximum value, \hessrev{while the maximum energy is determined by the last energy bin used in the spectrum, where the number of background events is $N_{\mathrm{OFF}}\geq 10$.}

\section{Multiwavelength observations}
For follow-up observations triggered by the GFU stream or the identification of potential counterparts in single alerts, we conducted multiwavelength analyses using data from the \textit{Fermi}-Large Area Telescope (LAT), the \icrevB{X-Ray Telescope on the Neil Gehrels Swift Observatory (\hessrevA{\swift{}}), and the Automatic Telescope for Optical Monitoring (ATOM)}. Our analysis of \textit{Fermi}-LAT data involved a binned likelihood analysis using \texttt{Fermipy} (v.1.2.0) \citep{Wood2017}. We utilized the \texttt{P8R3\_SOURCE} event class and corresponding instrument response functions\footnote{\url{https://fermi.gsfc.nasa.gov/ssc/data/analysis/documentation/Cicerone/Cicerone_LAT_IRFs/IRF_overview.html}}, selecting events in the energy range of 100 MeV to 300 GeV within 15$^\circ$ of the neutrino position and applying a zenith angle cut of 90$^\circ$. 
The \texttt{PowerLaw} model\footnote{\url{https://fermi.gsfc.nasa.gov/ssc/data/analysis/scitools/source_models.html}} was used for the \icrevA{gamma-ray} spectrum, with Galactic and isotropic templates for the background\footnote{\url{https://fermi.gsfc.nasa.gov/ssc/data/access/lat/BackgroundModels.html}}, and all catalog sources within 25$^\circ$ of the neutrino position were included~\citep{4FGL}. Each analysis covered a two-day time-scale, spanning a day before and after the \hessrevA{arrival time} of the neutrino events, in order to be sensitive to the detection of fast, bright transients, even down to durations of a few hours. No significant emission from the neutrino events was observed in the \textit{Fermi}-LAT data during the two-day time-scale that included the H.E.S.S. observations, resulting in the computation of 95\% C.L. upper limits assuming an $E^{-2}$ spectrum for the differential energy flux between 100 MeV and 300 GeV.

We conducted X-ray analysis using data from \icrevB{\hessrevA{\swift{}}} \citep{Burrows05}. The analysis was performed using \hessrevA{version 6.29} of the HEASOFT software~\cite{10.1111/j.1365-2966.2009.14913.x}, with data collected around the time of the neutrino ToO observations. The data utilized and the results of the analysis integrated in the 0.3-\icrevA{10}~keV energy range are presented in  \autoref{tab:swiftXRT}.

Optical data in the B, V, R, and I bands (440 nm, 550 nm, 640 nm, and 790 nm, respectively) were obtained using the 75 cm \icrevB{ATOM telescope} located at the H.E.S.S. site in Namibia \citep{ATOM}. The ATOM data analyzed in this study corresponds to observation times that coincide with the H.E.S.S. ToO observations. Energy fluxes \icrevA{for candidate sources} were derived for four neutrino follow-ups.

\begin{table*}
\centering







\resizebox{\textwidth}{!}{%
\begin{tabular}{cccccc}
  \toprule
  \toprule
    \multirow{2}{*}{\sc Neutrino ToO} & \multirow{2}{*}{\sc Source} &  \multirow{2}{*}{\sc Obs. IDS} & {\sc Photon} & {\sc Energy Flux (unabs.)} & \sc{Galactic Column Density}\\ 
             &        &        &     \sc{Index}         & $\times 10^{-12}$ $\rm{erg}~\rm{cm}^{-2}~\rm{s}^{-1}$ &  $\times 10^{20}$ ~cm$^{-2}$\\  
        \midrule
     GFU PKS 0625-35 & PKS 0625-35 & 00039136003-00039136004 & $2.19^{0.19}_{-0.18}$ & $10.7^{+1.27}_{-1.00}$ & 7.9\\

     IC-211125A & 4FGL J0258.1+2030 & 00041549003 &  $1.60^{0.70}_{-0.50}$ & $1.80^{0.80}_{-0.50}$ & 24.0 \\
     
     GFU PKS 0829+046 & PKS 0829+046 & 00038144014 & $1.87^{0.33}_{-0.30}$ & $2.30^{0.40}_{-0.30}$ & 2.70 \\
     
     GFU 1ES 0229+200 &  1ES 0229+200 & 00031249115 & $1.83^{0.18}_{-0.17}$ & $17.96^{0.16}_{-0.15}$ & 12 \\
     
   \bottomrule
   \bottomrule
\end{tabular}}
\caption{\hessrevA{\swift{}} data analysis parameters. The unabsorbed energy flux is obtained by integrating over the full energy range 0.3-\icrevA{10}~keV, \hessrevA{correcting for the Galactic column density value~\citep{gal_columndensity}.}} 
\label{tab:swiftXRT}
\end{table*} 

\section{Results}

\subsection{GFU PKS 0625-35}
\begin{figure}
    \centering
     \includegraphics[width=0.7\textwidth]{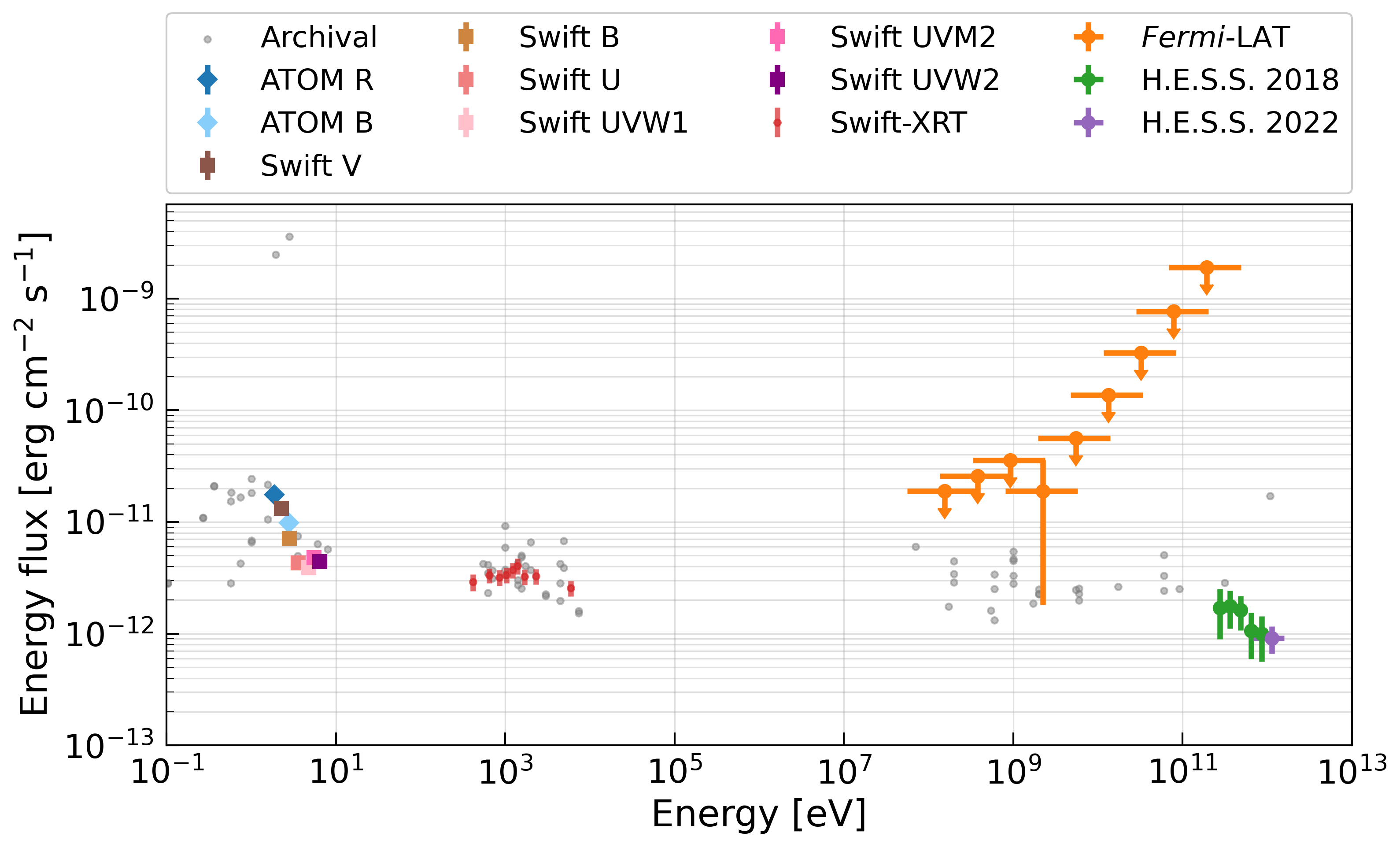}
    \caption{SED of the blazar PKS 0625-35 using data from \hess{}, \fermi{}, \hessrevA{\swift{}}, and ATOM. Archival data is displayed in gray.}
    \label{fig:pks0625_sed}
\end{figure}

On April 16, 2022, IceCube detected a neutrino flare originating from the blazar PKS 0625-35. The detection \icrevB{pre-trial} significance was $3.56\sigma$, with a false alarm rate of 0.003 per year. The three neutrinos were detected within an energy range of 63 TeV to 302 TeV. Given that the source had been previously observed by \hess{} \cite{pks0625}, a ToO observation was initiated on the night of April 19th, 2022 centered on the coordinates of the previous detection: RA: 6h26m58s, Dec: -35d29m50s. The observations lasted for three nights, resulting in approximately three hours of data. 
The source was detected in both \hess{} chain analyses, with a significance of $3.5\sigma$. The data was fitted with a power-law function, yielding a spectral index of $2.45 \pm 0.42$ in the energy range 0.35 to 6.58~TeV. The resulting spectral point is illustrated in \autoref{fig:pks0625_sed}, alongside the \hess{} flux points from the source's previous observation in 2018. During the same ToO, the source was also observed in optical and X-ray wavelengths. 
By comparing the flux distribution in April 2022 with archival data\footnote{\url{https://tools.ssdc.asi.it/SED/}} and the \hess{} data from 2018, it becomes evident that there was no variation in the shape of the \hessrevA{spectral energy distribution (SED)}, implying no association with the neutrino flare.

\subsection{IC-211125A}

\begin{figure}
     \centering
     \begin{subfigure}[b]{0.4\textwidth}
         \centering        \includegraphics[width=1\textwidth]{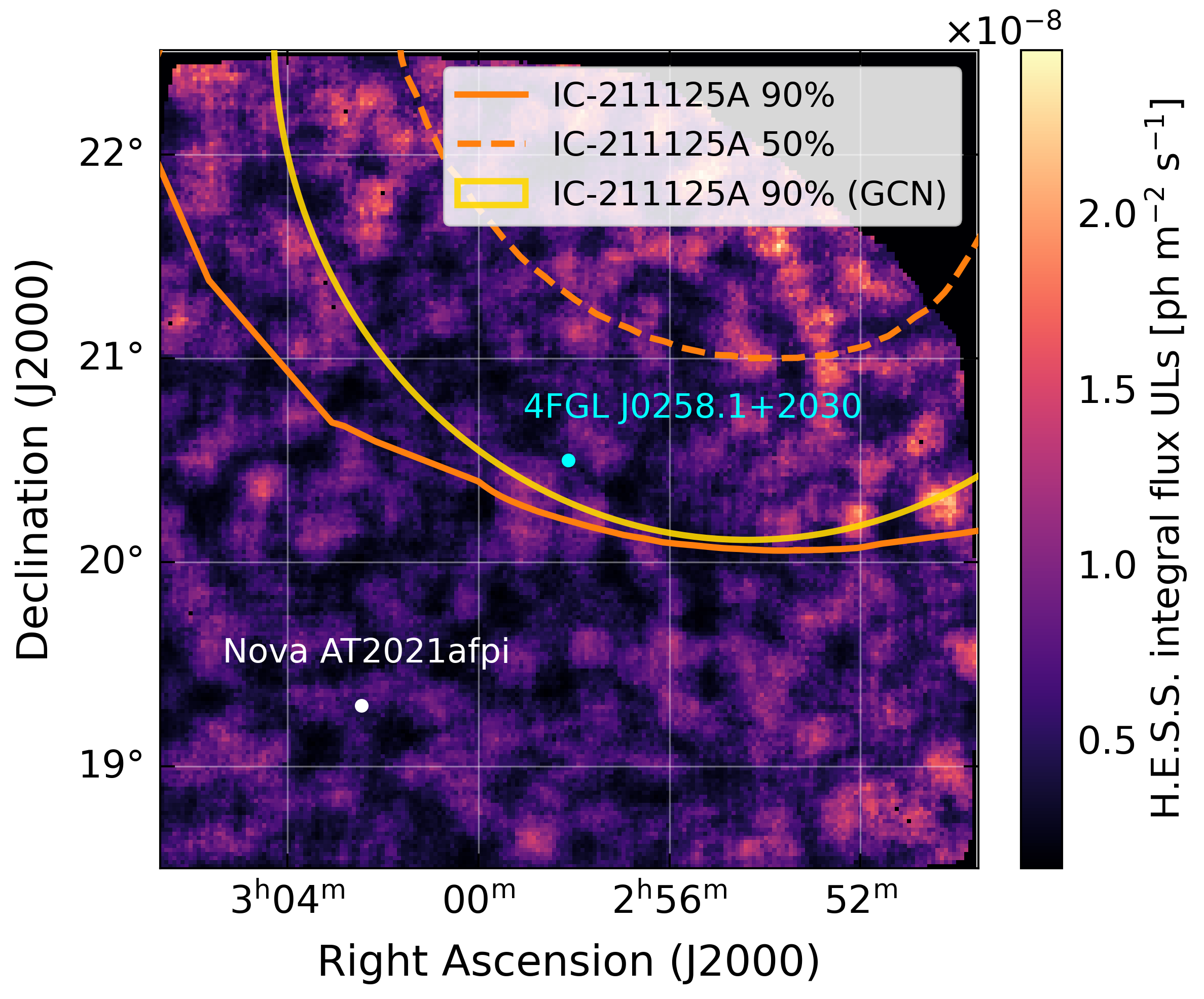}
        \caption{IC-211125A UL map}
        \label{fig:ic211125Amap}
     \end{subfigure}%
     \hfill
     \begin{subfigure}[b]{0.6\textwidth}
         \centering
         \includegraphics[width=0.9\textwidth]{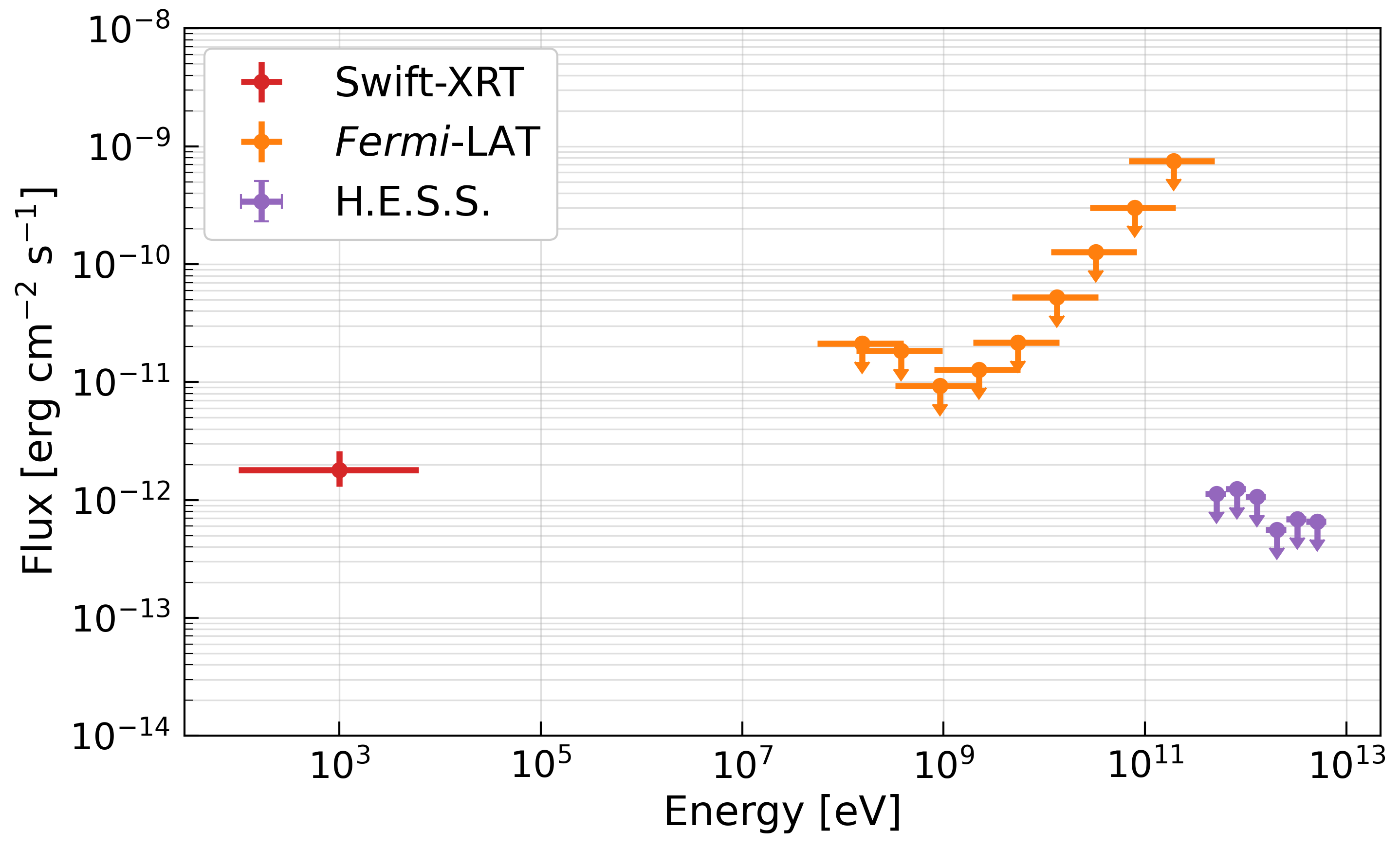}
        \caption{IC-211125A SED}
        \label{fig:ic211125Ased}
     \end{subfigure}%
    \caption{\textbf{(a)}: IC-211125A integral upper limits map of H.E.S.S. observations at 95\% C.L.  \textbf{(b)}: SED \icrevA{of the candidate source 4FGL J0258.1+2030} with \fermi{} upper limits and \hessrevA{\swift{}} data.}
    \label{fig:ic211125A}
\end{figure}

On November 25, 2021, IceCube reported the detection of a \textit{bronze} track-like event with a 39\% probability of \hessrevA{having an} astrophysical origin and a false alarm rate of 1.973 events per year due to atmospheric backgrounds~\cite{ic211125A}. Two sources in the ROI were identified as potential counterparts to the neutrino event: the bright optical transient AT2021afpi, which was detected by MASTER~\cite{master_atel} and classified as a classical nova~\cite{2021ATel15069}, and the AGN \fermi{} source 4FGL J0258.1+2030.
During the first night of observations on November 29th, \hessrevA{2021} \hess{} pointed at the position of the Nova source with a \hessrevA{0.5$^{\circ}$ ``wobble" offset\footnote{In wobble observation mode the source direction has an offset in declination with respect to the camera center.}}. In the following four nights, a manual pointing pattern without wobble offsets was defined to cover both sources. However, due to bad weather conditions and resulting unstable trigger rates, only $\sim5$~hours of observations were analyzed. \hessrevB{Upper limits were then computed} in the energy range between 0.42 and 100~TeV centered on the AGN position, as shown in \autoref{fig:ic211125Amap}. The yellow circle represents the first 90\% error on the neutrino position, while the updated contours provided by IceCube are depicted as orange circles. \hessrevB{The updated contours are calculated offline based on more sophisticated reconstruction algorithms.} Notably, the Nova source lies outside the 90\% IceCube contours. The minimum energy at the Nova position is 0.46~TeV.
Additionally, both \fermi{} and \hessrevA{\swift{}} conducted \hessrevA{observations during the neutrino alert}, and a \hessrevA{SED} was generated on the 4FGL J0258.1+2030 source position \icrevB{(see \autoref{fig:ic211125Ased})}.

\subsection{ IC-211208A} 
\begin{figure}
     \centering
     \begin{subfigure}[b]{0.4\textwidth}
         \centering
         \includegraphics[width=1.\textwidth]{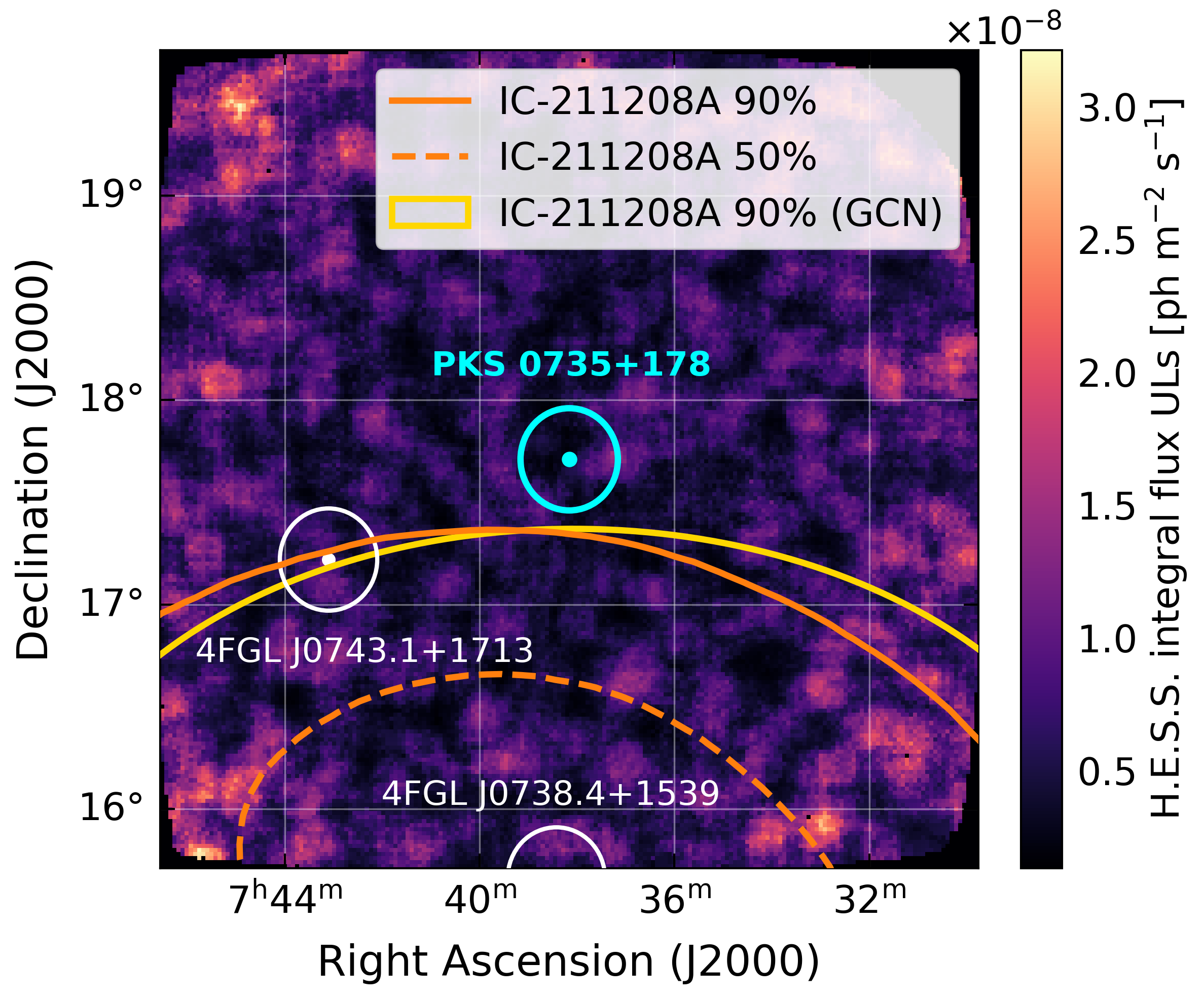}
        \caption{IC-211208A}
        \label{fig:ic211208Amap}
     \end{subfigure}%
     \hfill
     \begin{subfigure}[b]{0.6\textwidth}
         \centering
         \includegraphics[width=0.9\textwidth]{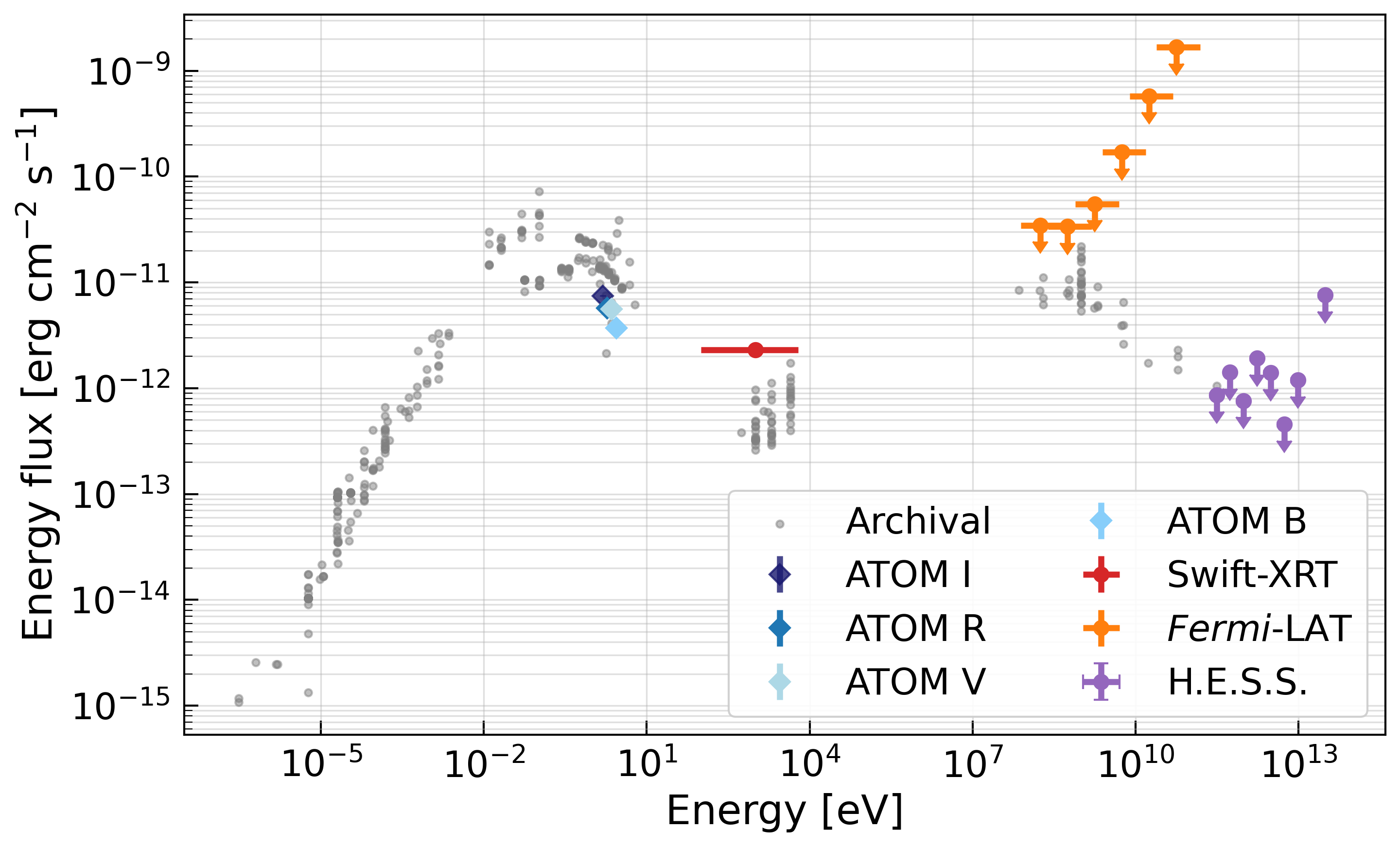}
        \caption{GFU PKS 0829+046}
        \label{fig:gfupks0829_sed}
     \end{subfigure}%
    \caption{\textbf{(a)}: IC-211208A integral upper limits map of H.E.S.S. observations at 95\% C.L. \icrevB{\textbf{(b)}:} SED of the PKS 0829+046 using data from \hess{}, \fermi{}, \hessrevA{\swift{}}, and ATOM. Archival data are displayed in gray.}
    \label{fig:ic211208A_gfupks0829}
\end{figure}

On December 8, 2021, IceCube reported the detection of a track-like neutrino event classified as a \textit{bronze} alert \hessrevA{with a 50.2\% probability of being astrophysical in} origin~\cite{2021GCN.31191....1I}. Three potential \hessrevB{AGN} gamma-ray counterparts were identified:  4FGL J0738.4+1539, 4FGL J0743.1+1713, and PKS 0735+178 (redshift $z=0.45$).  
Multiwavelength observations revealed that the gamma-ray blazar PKS~0735+178 was in flaring state in the radio, optical, X-ray, and GeV gamma-ray band~\citep{pks0735_icrc2023}.  As depicted in \autoref{fig:ic211208Amap}, PKS~0735+178 was located just outside the 90\% error region of the neutrino event, \hessrevA{$\sim 2.0^\circ$} away from the best-fit position. 
\hess{} observed the direction of PKS~0735+178 for a total of 16 hours from December 8 to 15, 2021, at an average zenith angle of $42.2^\circ$ \hessrevA{above 0.1~TeV energy}. \hessrevA{3.8} hours of data were selected based on strict criteria for weather conditions and instrumental status. Observations were performed in wobble mode at an offset from the center of the camera of $0.5^\circ$~\cite{hess_aharonian_2006}. A circular ROI of 0.1$^\circ$ centered on the position of PKS~0735+178 was defined, and two regions of 0.25$^\circ$ radius around  the nearby sources 4FGL~J0738.4+1539 and 4FGL~J0743.1+1713 were excluded from the background estimation. 
No significant gamma-ray excess above the expected background was detected from the direction of PKS~0735+178. Therefore, integral flux upper limits at 95\% C.L. in the energy range between 0.4 TeV and 100 TeV in the full field of view (FoV) were calculated (see \autoref{fig:ic211208Amap}). A complete SED of this \icrevA{source} \hessrevB{during IC-211208A alert} is given in~\cite{pks0735_icrc2023}. 

\subsection{GFU PKS 0829+046}
PKS 0829+046 is a BL Lac type object located at RA: 08h 31m 48.88s, Dec: +04$^\circ$ 29' 39.09''. The neutrino flare occurred on December 22nd, 2021, lasting  $\sim$7.9~days. It had a \icrevB{pre-trial} significance just above the trigger threshold (3.03$\sigma$) and a false alert rate of 0.055/year. The flare consisted of 8 events, with the most energetic event having an energy of 2~TeV.
Observations by \hess{} were conducted for two nights starting on December 30th, 2021, resulting in $\sim$4.7~hours of data. \hessrevB{In contrast to the other analyses, the \textit{loose cuts} configuration~\cite{hess_denaurois_2009} has been employed in this particular case.} However, the source was not detected at VHE, and an upper limit on the integral flux was computed above an energy threshold of 224~GeV. Similarly, the \fermi{} telescope did not detect the source. Nevertheless, \hessrevB{during the IceCube alert}, the source was detected in the optical band by ATOM and in the X-ray band by \hessrevA{\swift{}}. The flux and upper limit data points are presented in \autoref{fig:gfupks0829_sed}. Comparisons with archival data reveal no significant variation in the SED associated with the neutrino flare.

\subsection{IC-220425A}
\begin{figure}
     \centering
     \begin{subfigure}[b]{0.4\textwidth}
         \centering
         \includegraphics[width=1\textwidth]{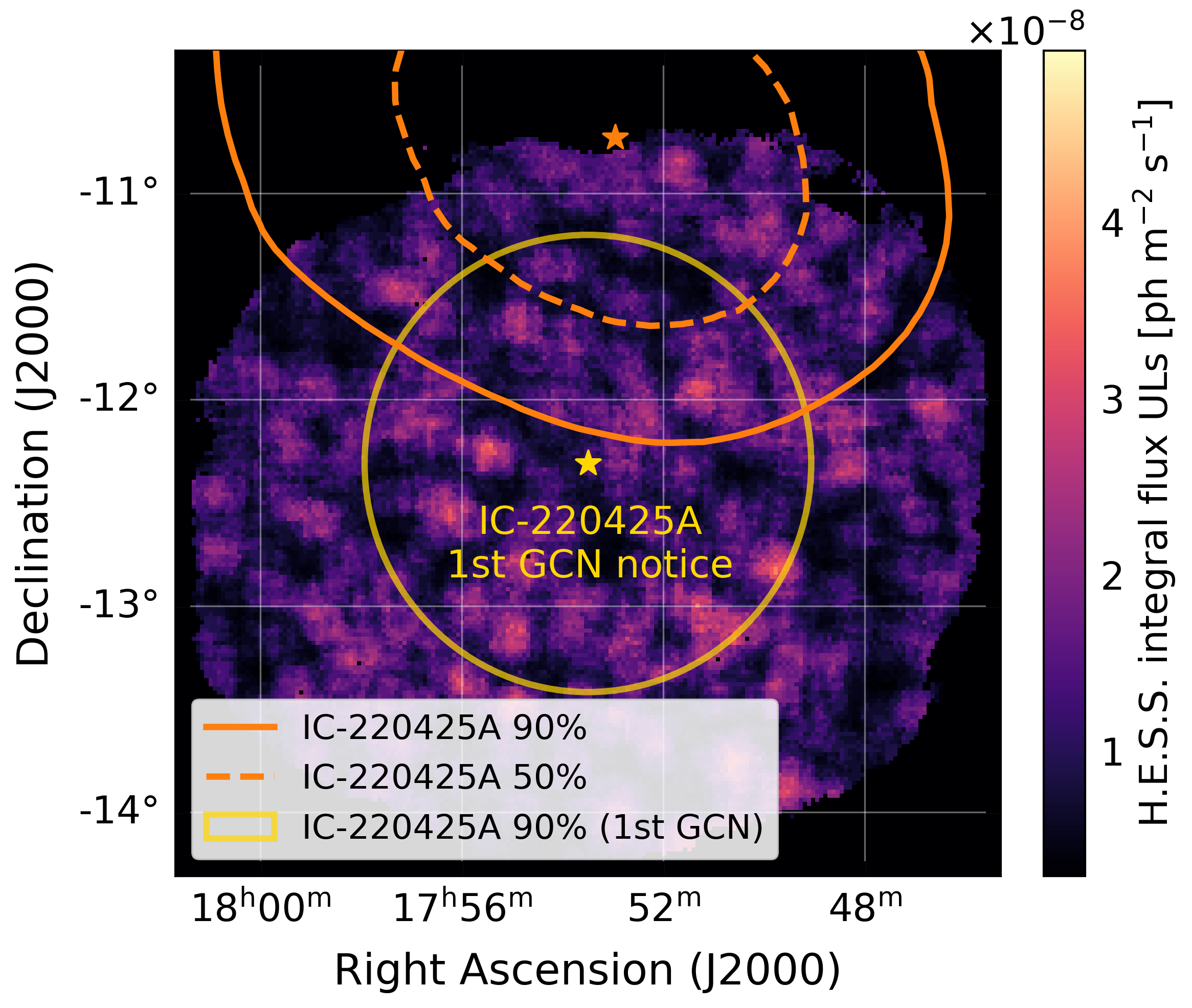}
        \caption{IC-220425A}
        \label{fig:ic220425A}
     \end{subfigure}%
     \hfill
     \begin{subfigure}[b]{0.6\textwidth}
         \centering
         \includegraphics[width=0.9\textwidth]{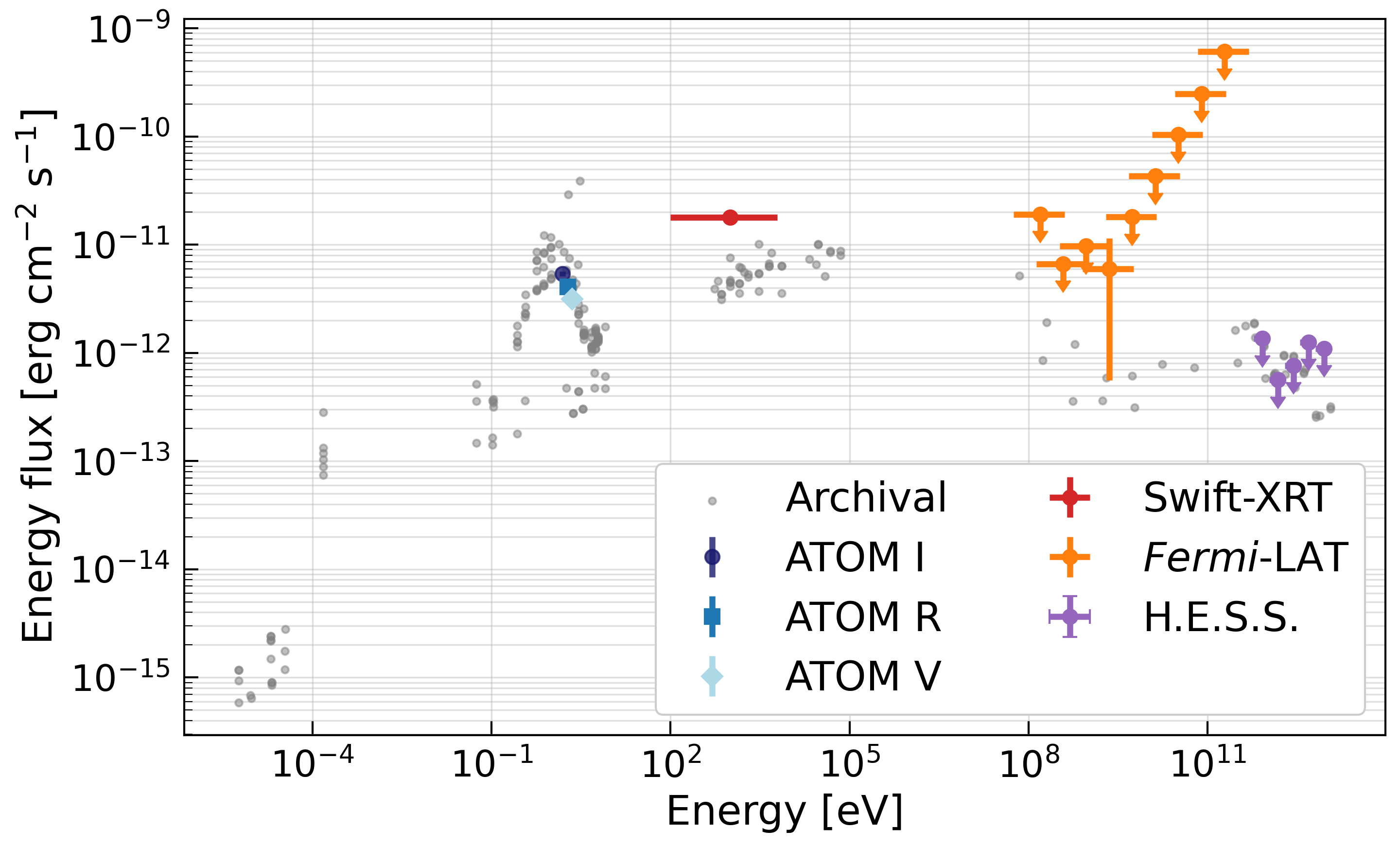}
        \caption{\icrevB{GFU 1ES0229+200}}
        \label{fig:gfu1es0229_sed}
     \end{subfigure}%
    \caption{\textbf{(a)}: IC-220425A integral upper limits map between 0.37 TeV and 100 TeV at a 95\% C.L. \textbf{(b)}: SED of the HBL \icrevB{GFU 1ES0229+200} using data from \hess{}, \fermi{}, \hessrevA{\swift{}}, and ATOM. Archival data is displayed in gray.}
    \label{fig:ic220425A_gfu1es0229}
\end{figure}

The \textit{gold} alert IC-20220425A was observed on April 25, 2022 and reported with a signalness of 17\%~\cite{ic220425A}. The \hess{} observation was triggered automatically within \hessrevA{less than 70 seconds} from the neutrino detection, resulting in 75 minutes observations \hessrevA{under} moonlight conditions. Due to lack of detection, 95\% C.L. integral upper limits were calculated between 0.37~TeV and 100~TeV and using \hessrevA{ad-hoc simulations of the instrument response} to take into account moonlight condition~\cite{rws_hess}. As depicted in \autoref{fig:ic220425A}, the updated neutrino position was refined offline, causing the \hess{} observation to fall outside the 90\% neutrino error region. The \fermi{} analysis revealed no evidence of potential high-energy gamma-ray counterparts.

\subsection{GFU 1ES0229+200}
1ES0229+200 is a high-frequency peaked BL Lac (HBL) located at RA: 02h 32m 49s, Dec: +20d 17m 1s~\cite{hess_1es0229}. On August 25th, 2022, IceCube detected a short neutrino flare from this source, lasting 1.2 days. The flare had a \icrevB{pre-trial} significance of $3.09\sigma$ and a false alert rate of 0.050/year. Only 4 neutrino events were detected, with the most energetic event having an energy of 1.3 TeV.
Observations by \hess{} were carried out over five nights starting on August 27th, 2022, resulting in approximately 13.5 hours of data. However, the \icrevA{observations} were affected by unstable weather conditions and only $\sim$8~hours are passing quality criteria. During the \hessrevB{neutrino alert}, the source was observed in the optical band by ATOM, in the X-ray band by \hessrevA{\swift{}}, and in the high-energy gamma-ray band by \fermi{}. The flux and upper limit data points are presented in \autoref{fig:gfu1es0229_sed}. Comparisons with archival data reveal no significant variation in the spectral distribution associated with the neutrino flare.

\section{Conclusions}

We presented the follow-up strategy of H.E.S.S. to real-time neutrino alerts and summarized \hessrevA{the search for gamma-ray emission} from November 2021 to the end of 2022.  While no significant gamma-ray excess was detected in the specific events studied, except for the blazar PKS~0625-35, the program contributed valuable insights into the search for neutrino sources and demonstrated the importance of multi-messenger observations in understanding high-energy astrophysical phenomena.

\bibliographystyle{ICRC}
\bibliography{references}

\clearpage

\section*{Full Author List: H.E.S.S. Collaboration}

\scriptsize
\noindent
F.~Aharonian$^{1,2,3}$, 
F.~Ait~Benkhali$^{4}$, 
J.~Aschersleben$^{5}$, 
H.~Ashkar$^{6}$, 
M.~Backes$^{7,8}$, 
V.~Barbosa~Martins$^{9}$, 
R.~Batzofin$^{10}$, 
Y.~Becherini$^{11,12}$, 
D.~Berge$^{9,13}$, 
K.~Bernl\"ohr$^{2}$
% ...

\medskip

\noindent
$^{1}$Dublin Institute for Advanced Studies, 31 Fitzwilliam Place, Dublin 2, Ireland\\
$^{2}$Max-Planck-Institut f\"ur Kernphysik, P.O. Box 103980, D 69029 Heidelberg, Germany\\
$^{3}$Yerevan State University,  1 Alek Manukyan St, Yerevan 0025, Armenia\\
$^{4}$Landessternwarte, Universit\"at Heidelberg, K\"onigstuhl, D 69117 Heidelberg, Germany\\
$^{5}$Kapteyn Astronomical Institute, University of Groningen, Landleven 12, 9747 AD Groningen, The Netherlands\\
$^{6}$Laboratoire Leprince-Ringuet, École Polytechnique, CNRS, Institut Polytechnique de Paris, F-91128 Palaiseau, France\\
$^{7}$University of Namibia, Department of Physics, Private Bag 13301, Windhoek 10005, Namibia\\
$^{8}$Centre for Space Research, North-West University, Potchefstroom 2520, South Africa\\
$^{9}$DESY, D-15738 Zeuthen, Germany\\
$^{10}$Institut f\"ur Physik und Astronomie, Universit\"at Potsdam,  Karl-Liebknecht-Strasse 24/25, D 14476 Potsdam, Germany\\
$^{11}$Université de Paris, CNRS, Astroparticule et Cosmologie, F-75013 Paris, France\\
$^{12}$Department of Physics and Electrical Engineering, Linnaeus University,  351 95 V\"axj\"o, Sweden\\
$^{13}$Institut f\"ur Physik, Humboldt-Universit\"at zu Berlin, Newtonstr. 15, D 12489 Berlin, Germany\\
% ...

\subsection*{Acknowledgements}

\noindent
The support of the Namibian authorities and of the University of
Namibia in facilitating the construction and operation of H.E.S.S.
is gratefully acknowledged, as is the support by the German
Ministry for Education and Research (BMBF), the Max Planck Society,
the Helmholtz Association, the French Ministry of
Higher Education, Research and Innovation, the Centre National de
la Recherche Scientifique (CNRS/IN2P3 and CNRS/INSU), the
Commissariat à l’énergie atomique et aux énergies alternatives
(CEA), the U.K. Science and Technology Facilities Council (STFC),
the Irish Research Council (IRC) and the Science Foundation Ireland
(SFI), the Polish Ministry of Education and Science, agreement no.
2021/WK/06, the South African Department of Science and Innovation and
National Research Foundation, the University of Namibia, the National
Commission on Research, Science \& Technology of Namibia (NCRST),
the Austrian Federal Ministry of Education, Science and Research
and the Austrian Science Fund (FWF), the Australian Research
Council (ARC), the Japan Society for the Promotion of Science, the
University of Amsterdam and the Science Committee of Armenia grant
21AG-1C085. We appreciate the excellent work of the technical
support staff in Berlin, Zeuthen, Heidelberg, Palaiseau, Paris,
Saclay, Tübingen and in Namibia in the construction and operation
of the equipment. This work benefited from services provided by the
H.E.S.S. Virtual Organisation, supported by the national resource
providers of the EGI Federation.
\clearpage
\section*{Full Author List: IceCube Collaboration}

\scriptsize
\noindent
R. Abbasi$^{17}$,
M. Ackermann$^{63}$,
J. Adams$^{18}$,
S. K. Agarwalla$^{40,\: 64}$,
J. A. Aguilar$^{12}$,
M. Ahlers$^{22}$,
J.M. Alameddine$^{23}$,
N. M. Amin$^{44}$,
K. Andeen$^{42}$,
G. Anton$^{26}$
% ...
\\
\\
% ...
$^{12}$ Universit{\'e} Libre de Bruxelles, Science Faculty CP230, B-1050 Brussels, Belgium \\
% ...
$^{17}$ Department of Physics, Loyola University Chicago, Chicago, IL 60660, USA \\
$^{18}$ Dept. of Physics and Astronomy, University of Canterbury, Private Bag 4800, Christchurch, New Zealand \\
% ...
$^{22}$ Niels Bohr Institute, University of Copenhagen, DK-2100 Copenhagen, Denmark \\
$^{23}$ Dept. of Physics, TU Dortmund University, D-44221 Dortmund, Germany \\
% ...
$^{26}$ Erlangen Centre for Astroparticle Physics, Friedrich-Alexander-Universit{\"a}t Erlangen-N{\"u}rnberg, D-91058 Erlangen, Germany \\
% ...
$^{40}$ Dept. of Physics and Wisconsin IceCube Particle Astrophysics Center, University of Wisconsin{\textendash}Madison, Madison, WI 53706, USA \\
% ...
$^{42}$ Department of Physics, Marquette University, Milwaukee, WI, 53201, USA \\
% ...
$^{44}$ Bartol Research Institute and Dept. of Physics and Astronomy, University of Delaware, Newark, DE 19716, USA \\
% ...
$^{63}$ Deutsches Elektronen-Synchrotron DESY, Platanenallee 6, 15738 Zeuthen, Germany  \\
$^{64}$ Institute of Physics, Sachivalaya Marg, Sainik School Post, Bhubaneswar 751005, India \\
% ...

\subsection*{Acknowledgements}

\noindent
The authors gratefully acknowledge the support from the following agencies and institutions:
USA {\textendash} U.S. National Science Foundation-Office of Polar Programs,
U.S. National Science Foundation-Physics Division,
U.S. National Science Foundation-EPSCoR,
Wisconsin Alumni Research Foundation,
Center for High Throughput Computing (CHTC) at the University of Wisconsin{\textendash}Madison,
Open Science Grid (OSG),
Advanced Cyberinfrastructure Coordination Ecosystem: Services {\&} Support (ACCESS),
Frontera computing project at the Texas Advanced Computing Center,
U.S. Department of Energy-National Energy Research Scientific Computing Center,
Particle astrophysics research computing center at the University of Maryland,
Institute for Cyber-Enabled Research at Michigan State University,
and Astroparticle physics computational facility at Marquette University;
Belgium {\textendash} Funds for Scientific Research (FRS-FNRS and FWO),
FWO Odysseus and Big Science programmes,
and Belgian Federal Science Policy Office (Belspo);
Germany {\textendash} Bundesministerium f{\"u}r Bildung und Forschung (BMBF),
Deutsche Forschungsgemeinschaft (DFG),
Helmholtz Alliance for Astroparticle Physics (HAP),
Initiative and Networking Fund of the Helmholtz Association,
Deutsches Elektronen Synchrotron (DESY),
and High Performance Computing cluster of the RWTH Aachen;
Sweden {\textendash} Swedish Research Council,
Swedish Polar Research Secretariat,
Swedish National Infrastructure for Computing (SNIC),
and Knut and Alice Wallenberg Foundation;
European Union {\textendash} EGI Advanced Computing for research;
Australia {\textendash} Australian Research Council;
Canada {\textendash} Natural Sciences and Engineering Research Council of Canada,
Calcul Qu{\'e}bec, Compute Ontario, Canada Foundation for Innovation, WestGrid, and Compute Canada;
Denmark {\textendash} Villum Fonden, Carlsberg Foundation, and European Commission;
New Zealand {\textendash} Marsden Fund;
Japan {\textendash} Japan Society for Promotion of Science (JSPS)
and Institute for Global Prominent Research (IGPR) of Chiba University;
Korea {\textendash} National Research Foundation of Korea (NRF);
Switzerland {\textendash} Swiss National Science Foundation (SNSF);
United Kingdom {\textendash} Department of Physics, University of Oxford.

\end{document}